\documentclass[article,preprint,floatfix]{revtex4-1}
\usepackage{textcomp}
\usepackage{graphicx}
\usepackage{amssymb}
\usepackage{epsfig}
\usepackage{ifsym}
\usepackage{wasysym}
\usepackage{latexsym}

\begin{document}

\title{Approaching magnetic ordering in graphene materials by FeCl$_3$ intercalation}

\author{Thomas Hardisty Bointon$^\dag$}
\author{Ivan Khrapach$^\dag$}
\author{Rositza Yakimova$^\ddag$}
\author{Andrey V. Shytov$^\dag$}
\author{Monica F. Craciun$^\dag$}
\author{Saverio Russo$^\dag$}
\email{Correspondence to S.Russo@exeter.ac.uk}
\affiliation{$^\dag$ Centre for Graphene Science, College of Engineering, Mathematics and Physical Sciences, University of Exeter, Exeter EX4 4QF, UK\\
$^\ddag$ Link\"oping University, S-58183 Link\"oping, Sweden}

\begin{abstract}
We show the successful intercalation of large area (1 cm$^2$) epitaxial few-layer graphene grown on 4H-SiC with FeCl$_3$. Upon intercalation the resistivity of this system drops from an average value of $\approx 200 \ \Omega/sq$ to $\approx 16 \ \Omega/sq$ at room temperature. The magneto-conductance shows a weak localization feature with a temperature dependence typical of graphene Dirac fermions demonstrating the decoupling into parallel hole gases of each carbon layer composing the FeCl$_3$ intercalated structure. The phase coherence length ($\approx 1.2 \mu$m at 280 mK) decreases rapidly only for temperatures higher than the 2-D magnetic ordering in the intercalant layer while it tends to saturate for temperatures lower than the antiferromagnetic ordering between the planes of FeCl$_3$ molecules providing the first evidence for magnetic ordering in the extreme two-dimensional limit of graphene.
 
\end{abstract}

\maketitle

The chemical functionalization of graphene with atomic species and/or molecules offers a unique way to engineer novel physical properties which are not found in pristine graphene \cite{Russo_review}. For example the functionalization with fluorine or hydrogen is known to open an energy-gap whose value is a function of the coverage of adatoms \cite{Martins2013,Withers2011b}. On the other hand, the intercalation of guest atoms or molecules into few layer graphene grants access to unique conductive \cite{Khrapach:2012cz}, magnetic \cite{PhilipKim:2011} and superconducting \cite{Kanetani:2012} properties of these materials. Recent experiments on intercalation with FeCl$_3$ of mechanically exfoliated few-layer graphene have reported an extraordinarily large charge doping (up to $9*10^{14} cm^{-2}$) with a record low electrical resistivity ($<8 \ \Omega/\rm{sq}$) and high optical transparency in the visible wavelength ($>$ 84 $\%$) \cite{Khrapach:2012cz}, which make this material attractive for applications requiring transparent electrodes. At the same time the experimental realization of Ca- and FeCl$_3$- intercalated bilayer graphene \cite{PhilipKim:2011,Kanetani:2012} provides a promising platform to investigate magnetism and superconductivity in two-dimensional systems.

So far most of the scientific effort has focused on the study of chemical functionalization of mechanically exfoliated graphene and of graphene grown by chemical vapour deposition, whereas the functionalization of epitaxially grown graphene on SiC is still in its infancy \cite{Riedl2009, Gierz2010}. Since epitaxially grown graphene on SiC is one of the main contenders for large area manufacturing \cite{Emtsev2009Graphene} of high quality graphene, understanding the chemical functionalization of epitaxial graphene is a fundamental first step to ensure the full exploitation of this material. Furthermore, SiC-based electronic devices are uniquely suited for complementary applications to those targeted by standard Si-based electronics \cite{Wright2008}. Therefore, functionalized epitaxial graphene on SiC could enable the development of a new class of electronic applications for hostile-environment working conditions -e.g. space applications with high radiation levels \cite{Godignon2011} or nuclear power centrals. From a fundamental point of view, functionalized epitaxial graphene could enable the study of states of matter such as magnetism and intrinsic superconductivity in the extreme two-dimensional limit inherent in this atomically thin material. 

Here we present a detailed study of FeCl$_3$ intercalated few-layer epitaxially grown graphene on 4H-SiC with a surface area of $\approx$ 1 cm$^2$, see Figure 1a. From the analysis of the Raman spectra and conductivity measurements we find that the intercalation process is uniform and the typical room temperature square resistance drops from $174 \pm 9 \ \Omega/sq$ to $16.6 \pm 0.6 \ \Omega/sq$ upon intercalation. The measured magneto-conductance (MC) shows a strong weak localization (WL) feature at cryogenic temperatures ($<25K$) whose temperature dependence is well-described by the WL theory of a single layer graphene \cite{McCann:2006ip}. These findings demonstrate that the intercalation of FeCl$_3$ originates parallel single layer-like hole gases in the stacking with a phase coherence length ($L_\phi$) as large as $1.17 \ \pm 0.08 \mu$m at 280 mK. The temperature dependence of $L_\phi$ shows a steep decrease of $L_\phi$ for temperatures higher than $\sim 30K$, a temperature compatible with 2-D magnetic ordering in the FeCl$_3$ layer, while $L_\phi$ tends to saturate at temperatures lower than $\sim 4K$, a temperature compatible with 3-D antiferommagnetic coupling between planes of FeCl$_3$. These observations show that FeCl$_3$ intercalated epitaxial graphene is a unique system to study magnetic ordering in the extreme two-dimensional limit of graphene. Finally the intercalation of FeCl$_3$ offers a simple way to induce high charge carrier density in graphene ($>1.5 \times 10^{14} \rm{cm}^{-2}$) which is pivotal for pioneering the physics of plasmons in unprecedented ranges of wavelengths \cite{Grigorenko2012}.

\begin{figure}{}
\center
\includegraphics[scale=0.45]{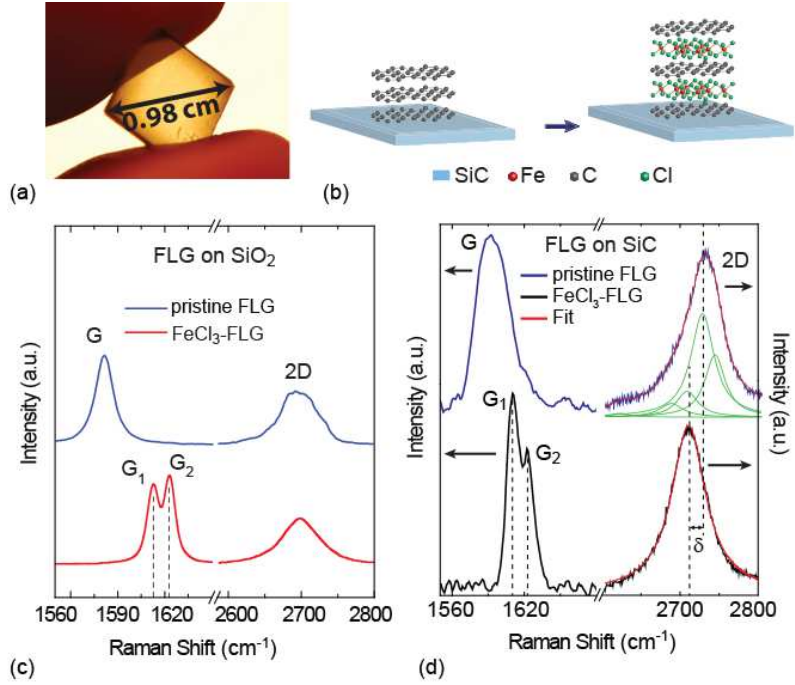}
\caption{\label{fig1} Panels (a) and (b) show a picture of the intercalated FLG on 4H-SiC and a schematic crystal structure of trilayer FeCl$_3$-FLG, respectively. The graph in (c) is a plot of the G and 2D Raman bands for mechanically exfoliated pristine (top) and FeCl$_3$ intercalated trilayer graphene (bottom) on SiO$_2$. d) Shows the Raman spectra for epitaxial pristine FLG (top) and FeCl$_3$ intercalated FLG (bottom). The red curve is a fit of the 2D peak to a multiple Lorentzian function (green curves). Upon intercalation with FeCl$_3$ a single Lorentzian is the best fit of the 2D peak (red).}
\end{figure}

Pristine few-layer graphene (FLG) was grown via thermal decomposition of the silicon terminated face of 4H-SiC \cite{Virojanadara2008,Yakimova2010}. The layer number of pristine graphene was determined by Raman spectroscopy. The intercalation of FLG with FeCl$_3$ is schematically shown in Figure 1b. This process was performed in vacuum for a total of 24 hours using the two zone vapour transport technique described by Khrapach \textsl{et al.} \cite{Khrapach:2012cz}. Raman spectroscopy and electrical transport measurements were employed to characterise the structure and macroscopic sheet resistance of the intercalated samples. To ascertain the intrinsic electrical properties of this material we measured the magneto-resistance with standard lock-in technique in Hall bar devices (length 200 $\mu$m and width 50 $\mu$m) from 200K down to 280mK and with perpendicular external magnetic fields up to $\pm$ 2T.

Recent experiments on FeCl$_3$ intercalated FLG obtained from mechanically exfoliated natural graphite have reported a large charge transfer between the intercalant and graphene \cite{Khrapach:2012cz}. This modifies the Raman spectra of FLG in two distinct ways: an upshift of the G-band and a change of the 2D-band from multi- to single-peak structure, respectively (see Figure 1c). A graphene sheet which has only one adjacent FeCl$_3$ layer gives rise to the G$_1$-band (1612cm$^{-1}$), whereas a graphene sheet sandwiched between two FeCl$_3$ layers is characterised by the G$_2$-band (1625cm$^{-1}$) \cite{Zhan:2010jf, Zhao:2011ut,Khrapach:2012cz}. The changes observed in the shape of the 2D-band indicate decoupling of the FLGs into separate monolayers due to the intercalation of FeCl$_3$ between the graphene sheets.

Figure 1d shows a comparison of the measured Raman spectra for pristine and intercalated FLGs. Unlike graphene deposited on SiO$_2$, the Raman spectra of epitaxial graphene contains the signal from the SiC substrate in the spectral region of the G band. To separate the 4H-SiC and graphene Raman spectra we have subtracted the spectrum of 4H-SiC in the G-band region by following the procedure introduced by Rohrl \textit{et al.} \cite{Rohrl:2008et}. For pristine FLG on 4H-SiC, both G and 2D bands are upshifted as compared to pristine FLG on SiO$_2$ due to the substrate induced strain \cite{Ni:2008du,Rohrl:2008et,Lee:2008fx}. To identify the number of layers in pristine FLG on SiC we used the method developed by Lee \textit{et al.}, where it was shown that the value of the full width at half maximum of the 2D band (FWHM(2D)) exhibits a linear relationship with the inverse number of layers (N): $\rm{FWHM(2D)=(-45(1/N)+88) [cm^{-1}]}$ \cite{Lee:2008fx}. In our epitaxial FLG samples, the 2D band has a FWHM of 72 cm$^{-1}$, from which we estimate a layer number of three (see Figure 1d). Upon intercalation with FeCl$_3$, the G-band of epitaxial FLG upshifts to the G$_1$ (1610 cm$^{-1}$) and G$_2$ (1622 cm$^{-1}$) positions demonstrating the successful intercalation of FeCl$_3$ molecules in the epitaxial FLG. At the same time, the 2D-band structure of the FLG on SiC is changed from a multi-peak structure to that of a single Lorentzian and the 2D-band downshifts to that of monolayer graphene on 4H-SiC ($\approx$2715cm$^{-1}$), see Figure 1d. These observations demonstrate that our FeCl$_3$-FLG on SiC has three graphene sheets and two FeCl$_3$ layers intercalated between the graphene sheets as schematically illustrated in Figure 1b. Therefore, this intercalation stage leads to the formation of 3 parallel hole gases, where the top and bottom gases will have similar values of charge density which will be lower than the charge density of the middle hole gas.

\begin{figure}{}
\center
\includegraphics[scale=0.8]{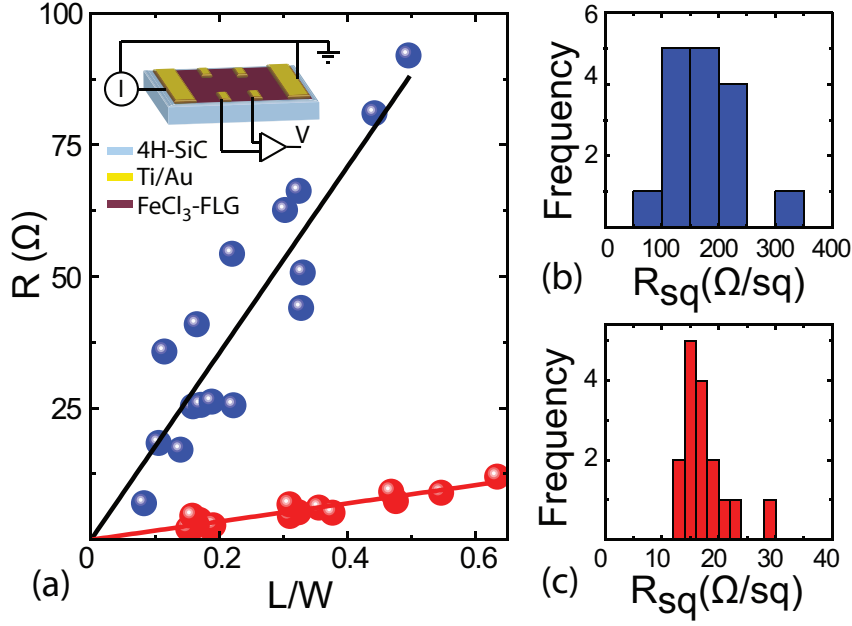}
\caption{\label{fig2} a) Shows a plot of the measured four terminal resistance (R) as a function of the ratio between the conductive channel lenght (L) and width (W). The blue data points correspond to pristine FLG on 4H-SiC while the red data points correspond to FeCl$_3$-FLG on 4H-SiC. The inset shows a scheme of the configuration used to measure the resistivity. b) and c) are histograms showing the distribution of resistivity for pristine FLG (Blue) and FeCl$_3$-FLG (Red), respectively.}
\end{figure}

The sheet resistance of the as grown epitaxial FLG and FeCl$_3$-intercalated epitaxial FLG was first measured in a four terminal configuration in macroscopic samples with a conductive channel of fixed width (0.7 cm) and channel length ranging from 0.7 mm to 4.2 mm using Ti/Au (5/50 nm) contacts. It was found experimentally that the pristine epitaxial FLG have an average sheet resistance ($R_{sq}$) of $174 \pm 9 \ \Omega$/sq, estimated from the linear fit shown in Figure 2a. The corresponding values of resistance in the same devices after intercalation with FeCl$_3$ are systematically much lower with an average value of $16.6 \pm 0.6 \ \Omega $/sq. The distribution of $R_{sq}$ measured for FLG and FeCl$_3$-FLG are presented as two histograms, see Figure 2b and c. In both cases we find a single peaked distribution with narrow spreading of the values of $R_{sq}$ demonstrating the homogeneous and reproducible electrical properties in large area FeCl$_3$-FLG. 

The charge carriers sign and density are readily characterized from measurements of the transverse magneto-resistivity ($\rho_{\rm{xy}}$) in Hall bar devices (see inset of Figure 3a). The positive slope of $\rho_{\rm{xy}}(\rm{B})$ stems for holes as the dominant charge carrier type with a typical density of $n_h \geq 1.5\times10^{14}$ cm$^{-2}$. This high charge carrier density far exceeds the values achievable with top-gates on SiC\cite{Wu:2008gj} and is comparable to that achieved in electric-double-layer FLG transistors \cite{Chen:2012fj,Ye2011}. 

Measurements of the temperature dependence of the longitudinal resistivity reveal that $\rho_{\rm{xx}}$ decreases monotonously from 175K down to 25K as expected for heavily doped graphene \cite{Khrapach:2012cz}, see Figure 3b. However, at temperatures below 25K we systematically observe an increase in $\rho_{\rm{xx}}$ which was not previously observed in intercalated mechanically exfoliated FLG and whose quantum nature can only be unveiled with a detailed study of the magnetic field dependence. Figure 3c shows a typical measurement of $\rho_{\rm{xx}}(\rm{B})$ at temperature T=300mK. A zero-field peak characteristic of weak localization is clearly visible \cite{Gorbachev:2007hb}. As the magnetic field breaks time-reversal symmetry, the quantum interference is reduced and a classical magneto-resistive signal is observed ($\left| \rm{B} \right| \geq \pm 250mT$). Note that owing to the macroscopic size of the epitaxial-FLG Hall bars we can directly measure the weak localization signal without the need for subtracting any mesoscopic conductance fluctuations typically present in devices fabricated on mechanical exfoliated graphene \cite{Gorbachev:2007hb}. 

\begin{figure}{}
\center
\includegraphics[scale=0.8]{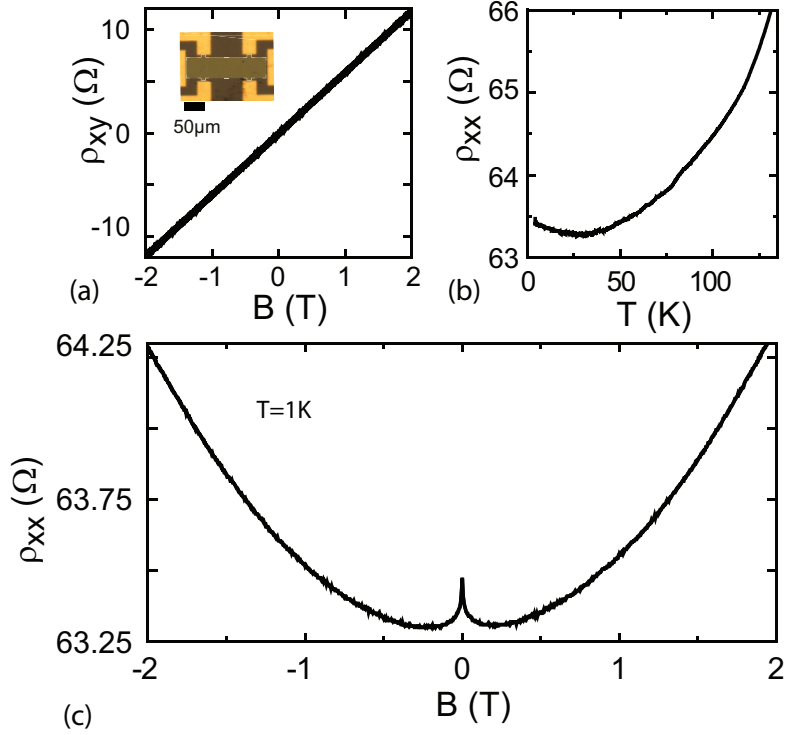}
\caption{\label{fig3} a) Shows a plot of the Hall resistivity ($\rho_{xy}$) measured in a representative Hall bar device shown in the micrograph picture in the inset. (b) and (c) show the temperature and perpendicular magnetic field dependence of the longitudinal resistivity ($\rho_{xx}$) measured in the same device.}
\end{figure}

A detailed study of the temperature dependence of the longitudinal magneto-resistivity shows that the WL peak is heavily suppressed when the temperature increases, see Figure 4a. At the same time the concavity of the roughly parabolic shape background of $\rho_{xx}$ \textit{vs.} B is clearly temperature independent. Recent experiments have shown that the longitudinal magneto-resistance can be understood as the sum of a Drude term plus a correction due to electron-electron interaction. The fingerprint of the electron-electron interaction correction is a strong temperature dependent concavity of $\rho_{\rm{xx}}(\rm{B})$ \cite{Jobst:2012ba}. The fact that our measurements show no significant change of the concavity as a function of the temperature (see Figure 4a) implies that for FeCl$_3$-FLG the electron-electron interaction correction to the magneto-resistance is negligible compared to the classical Drude term. Indeed, if we only consider the conductivity tensor with terms $\sigma_{\rm{xx,j}}=\frac{\sigma_{0,\rm{j}}}{1+(\mu_j B)^2}$ and $\sigma_{\rm{xy,j}}=\frac{\sigma_{0,\rm{j}} \mu_j B}{1+(\mu_j B)^2}$, where $\sigma_{0,\rm{j}}$ is the classical Drude conductivity for each hole gas indexed by $j$, we can describe accurately the experimental data using the equivalent magneto-resistance for 3 parallel hole gases of which two with similar charge density as previously identified from Raman spectroscopy, see Figure 4a. 

Further evidence that the hole gases present in FeCl$_3$-FLG are decoupled Dirac fermions is provided by a detailed analysis of the WL correction which we conduct on the signal after subtracting the classical magneto-resistance background, see Figure 4b. We consider the equivalent MC for the intercalated-FLG to be $\Delta \sigma = \sum ^3_{j=1} \Delta \sigma_j$, with $\Delta \sigma_j$ the well-established MC for Dirac fermions in single-layer graphene \cite{McCann:2006ip}:


\begin{equation}
\Delta\sigma_j(B) = \frac{e^2}{\pi h} \left(F\left(\frac{B}{B_{\phi,j}}\right)-F\left(\frac{B}{B_{\phi,j} + 2B_{i,j}} \right) -2F \left(\frac{B}{B_{\phi,j} + B_{i,j} + B_{\star,j}} \right) \right), 
\end{equation}

$B$ is the applied perpendicular magnetic field, $B_{\phi,i,\star} = \frac{\hbar }{4De}\tau^{-1}_{\phi,i,\star}$ and diffusion coefficient $D = \hbar v_F\sqrt{\pi}\slash 2e^2\rho_{xx}\sqrt{n}$, $F(z) = lnz + \psi\left(\frac{1}{2} +\frac{1}{z}\right)$ and $\psi$ is the digamma function. 

In graphene WL is uniquely sensitive to inelastic phase breaking processes characterized by a dephasing scattering time $\tau_{\phi}$ and also to elastic scattering that breaks the chirality of the charge carriers (e.g. scattering off surface ripples, dislocations and atomically sharp defects) with a corresponding scattering time $\tau_s$. A potential anisotropy of the Fermi surface in $k$ space, i.e. trigonal warping, can also destroy intravalley WL and the associated scattering time is $\tau_w$, and $\tau^{-1}_\star=\tau^{-1}_s+\tau^{-1}_w$. Finally the intervalley scattering which occurs at a rate $\tau^{-1}_i$ on defects with size of the order of the lattice spacing, promotes the interference of carriers from different valleys in momentum space. 
 
\begin{figure}{}
\center
\includegraphics[scale=0.5]{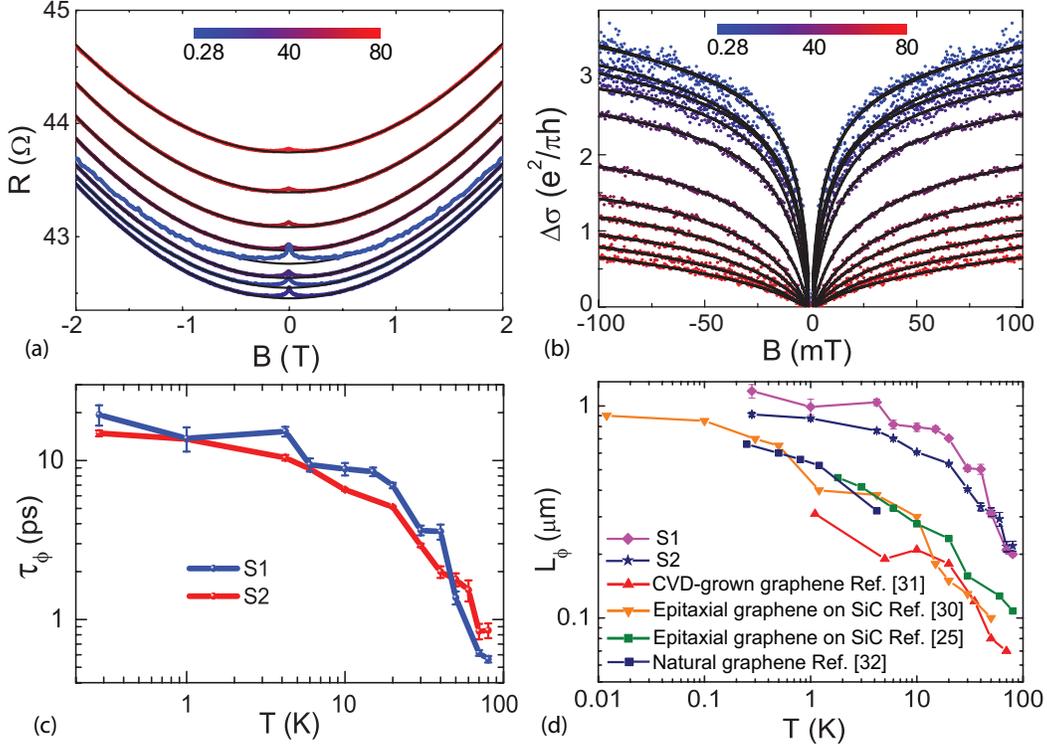}
\caption{\label{fig4} The graph in (a) is a plot of the measured magneto resistance (scatter points) at various temperatures (see colour coded legend). The continuous lines are a fit to the Drude conductivity. b) Shows the measured weak localization contribution to the magneto-conductance (scatter points) and the best fit (solid line) to the model of Eq. (1) for various temperatures. c) shows the temperature dependence for the extracted $\tau_{\phi}$ for two different intercalated samples (labelled S1 and S2) and (d) shows a summary of the temperature dependence of the values of $L_{\phi}$ present in literature relative to pristine single-layer graphene grown and/or obtained by different methods as specificed in the legend and a comparison to the values of $L_{\phi}$ estimated in our FeCl$_3$-FLG.}
\end{figure}

Figure 4b shows the measured WL signal extracted from the longitudinal magneto-resistivity after subtracting the classical background and the corresponding best fit of the total MC modelled with the parallel of three decoupled hole gases for different temperatures. At T=280mK the best-fit values for the middle hole gas with higher charge density are $B_i=7.4 * 10^{-4} \pm 6 * 10^{-5}$ T and $B_\phi= 1.2 * 10^{-4} \pm 2 * 10^{-5}$ T and for the outer hole gases with lower charge density $B_i = 4.6* 10^{-3} \pm 5 * 10^{-3}$ T and $B_\phi = 1.6 * 10^{-3} \pm 3*10^{-4}$. In this fit procedure we find that the term in Eq. (1) containing $B_\star$ is negligible, stemming for a very short $\tau_\star$ compared to the other characteristic times (i.e. $\tau_i$ and $\tau_\phi$). The good agreement between the experimental data and the model holds for a very wide range of temperatures starting from 280mK up to the highest measured (80K). We can therefore conclude that our samples are in the regime where $\tau_{tr} \approx \tau_\star<<\tau_i<<\tau_\phi$. If we assume an identical diffusion constant for each hole gas we can estimate the value of $\tau_\phi$ at each different measured temperature. Since the typical behaviour of the dephasing time, as predicted by theory is power-like \cite{AA} (e.g. $\tau_\phi \propto T^{-2/3}$ for narrow wires), we plot our data in a double logharitmic scale, see Figure 4c. This analysis reveals two different regimes: rapid decrease of the dephasing time for temperatures $T>4K$ and nearly saturated behaviour at low temperature. Such saturation is common for samples with magnetic impurities in microscopic concentrations \cite{Birge,Pierre2003} which in graphene can originate from chemisorbed magnetic atoms, hydroxyl groups, vacancies or resonant localized states in the vicinity of the Dirac point \cite{Kozikov2012}. In our case the presence of iron in the intercalant specie is a source of magnetic impurities, which can account for the observed saturation of the dephasing time. 

The dephasing length can be directly extracted from the parameter $B_\phi$ using the relation $L_\phi=\sqrt{D \tau_\phi}$ which gives $B_\phi = \frac{\hbar}{4 e L^2_\phi}$. Therefore without making any assumption on the diffusion constant of the hole gases we extract the corresponding values of $L_\phi$ for a wide range of temperature (from 80K down to 280mK), see Figure 4d. Upon lowering the temperature the values of $L_\phi$ estimated for FeCl$_3$-epitaxial FLG increase monotonically, however below $\sim 4K$ $L_\phi$ this increase is less pronounced with a maximum value of $L_\phi = 1.17 \pm 0.08 \mu$m at 280mK. To understand the possible physical origin of the temperature dependence of $L_\phi$ in our material, we compare $L_\phi(T)$ for two representative FeCL$_3$-epitaxial FLG (i.e. indicated by S1 and S2) with values measured in pristine epitaxial FLG \cite{Lara-Avila2011,Jobst:2012ba}, CVD graphene \cite{Cao2010} and graphene mechanically exfoliated from natural graphite \cite{Tikhonenko2008}. The overall temperature dependence of $L_\phi$ is similar for all the different kinds of graphene, however two striking features distinguish FeCl$_3$-epitaxial FLG from pristine-FLG. More specifically FeCl$_3$-FLG has (1) larger values of $L_\phi$ than pristine graphene at a given temperature and (2) $L_\phi$ decreases rapidly only for temperatures much higher than what is observed in pristine graphene, see Figure 4d.

To explain the macroscopic scales of the quantum coherence estimated in our samples even in the presence of magnetic FeCl$_3$ molecules - expected to be a source of excess dephasing- and the temperature dependence of $L_\phi$ we have to consider what is known on the magnetic ordering in stage one FeCl$_3$-intercalated graphite. This material is a Heisenberg system characterized by two distinct magnetic orderings \cite{Corson1982,Simon1983,Dresselhaus}, these are (1) strong 2-D magnetic correlations in the plane of FeCl$_3$ with a transition temperature of $\sim 30$K and (2) a transition to 3-D antiferromagnetism between planes of FeCl$_3$ at 3.8K. These magnetic phases have been studied experimentally in bulk materials with M\"ossbauer and neutron diffraction \cite{Corson1982,Simon1983}. A theoretical understanding of the structural, electronic and magnetic properties in FeCl$_3$-FLG was only recently presented using \textit{ab initio} numerical simulations using a generalized gradient approximation of Perdew-Burke-Ernzerhof with an effective interaction parameter which accounts for the Hubbard parameter as well as the exchange interaction \cite{Li2013}. It was found that for stage one FeCl$_3$-FLG within the spin-polarized calculations at low temperature the 3-D magnetic ground state is antiferromagnetic, that is 1.5 eV per unit cell lower than the ferromagnetic configuration. At the same time the magnetic moments of FeCl$_3$ are expected to align ferromagnetically in the same intercalant layer. Finally, the band structure of this compound is also expected theoretically to have a linear dispersion with a Fermi energy of $\sim 0.9$eV \cite{Li2013,Zhan:2010jf} as confirmed by recent experimental studies \cite{Khrapach:2012cz}.    

Spin-flip scattering off time dependent randomly oriented magnetic moments is a cause of excess dephasing in mesoscopic systems \cite{Birge,Pierre2003,Kozikov2012}. However, upon attaining a macroscopic magnetic ordering, the charge carriers experience less spin-flip scattering events leading to a longer phase coherence length. Although the material studied in this article is not bulk, i.e. it is only a 3-layer FLG, we observe experimentally large values of $L_\phi$ for $T<30K$ (compatible with the 2-D magnetic correlations in the plane of FeCl$_3$) and a saturation of $L_\phi$ for T lower than $\sim 4$K, see Figure 4d. Therefore, considering what is known on the magnetic ordering in this kind of intercalated compounds \cite{Corson1982, Simon1983,Dresselhaus,Li2013}, our experimental findings strongly suggest that magnetic ordering of the FeCl$_3$ planes suppress magnetic scattering in FeCl$_3$-FLG. For temperatures higher than the 30K a sharper decrease of $L_\phi$ is observed in FeCl$_3$-FLG as compared to pristine graphene indicating that randomly oriented magnetic moments in the intercalated FLG are driving excessive dephasing.     

In conclusion we have demonstrated the successful intercalation of large area epitaxial graphene grown on 4H-SiC with FeCl$_3$. We found that this material has a very low square resistance ($\approx16 \Omega/\rm{sq}$) becoming an ideal candidate for transparent electrodes in novel complementary applications targeted by SiC-based electronics. With a detailed study of the Raman spectra we have characterized the structure of this material. Finally we have conducted a systematic study of MC in Hall-bar devices and we have estimated a large dephasing length of $L_\phi \approx 1.2 \mu $m at 280mK in FeCl$_3$-FLG. The observed temperature dependence of $L_\phi$ is compatible with the occurrence of magnetic ordering in stage one FeCl$_3$-FLG, showing that this novel material is a good platform for studying magnetic ordering in the extreme two-dimensional limit of graphene.

\noindent \textbf{Acknowledgements}
SR and MFC acknowledge financial support from EPSRC (Grant no. EP/J000396/1, EP/K017160/1, EP/K010050/1, EP/G036101/1). RY acknowledges financial support from the Swedish NRC, contract VR 324-2011-4447.

\end{document}